\documentclass [12pt]{article}
\usepackage {graphicx}
\usepackage {longtable}

\begin{document}

\bigskip

\section*{On the Possibility to Explain ``The Pioneer Anomaly'' within the Framework 
of Conformal Geometrodynamics}

\bigskip

\begin{center}
\textbf{M.V. Gorbatenko}
\end{center}

\bigskip

\begin{center}
Russian Federal Nuclear Center - All-Russian Research Institute of 
Experimental Physics, Sarov,
\end{center}

\begin{center}
Nizhni Novgorod region; E-mail: \underline {gorbatenko@vniief.ru}
\end{center}

\bigskip

\textbf{Abstract}

\bigskip

Einstein-Infeld-Hoffmann method is used to solve the problem of motion of 
two bodies when the equations of general relativity are of the generalized 
form: they have been reduced to a form invariant under conformal 
transformations. It is proved that not only metric degrees of freedom, but 
also derivatives of vector $A_{\alpha}  $ appearing in the generalized 
equations can exert influence on the motion of bodies in a certain 
space-time domain. This influence can account for the recently observed 
anomalous acceleration of spacecrafts Pioneer 10, Pioneer 11. The impact of 
vector $A_{\alpha}  $ on the motion of bodies is interpreted as a 
consequence of viscosity in geometrodynamic continuum.

\bigskip

Key words: EIH method; Pioneer 10, 11.

\bigskip

\subsection*{1. Introduction}

\bigskip

The analysis of the paths of spacecrafts Pioneer 10, Pioneer 11 in refs. 
[1], [2] suggests that in range

\begin{equation}
\label{eq1}
R = \left( {30 \div 100} \right)AU = \left( {0.45 \div 1.5} \right) \cdot 
10^{15}\;cm
\end{equation}

\noindent
the spacecrafts are subject to an anomalous acceleration component, which 
has come to be denoted by $a_{P} $ in the literature. In magnitude the 
acceleration $a_{P} $ is 
\begin{equation}
\label{eq2}
a_{P} = - \left( {8.74 \pm 1.33} \right) \cdot 10^{ - 8}\;cm/s^{2}.
\end{equation}
The minus sign indicates that the direction of the acceleration $a_{P} $ is 
close to the direction towards the Sun.

From the viewpoint of the following discussion of the issue of the 
acceleration $a_{P} $ it seems reasonable to make two estimations. First, we 
should compare (\ref{eq2}) to acceleration $a_{N} $ gained by spacecrafts from the 
Sun in the Newtonian approximation. Having taken the mass of the Sun equal 
to $M = 2 \cdot 10^{33}\;g$, gravitational constant $G = 6.67 \cdot 10^{ - 
8}\;cm^{3}/g \cdot s^{2}$, we obtain for distance $R = 50\;AU = 7.5 \cdot 
10^{14}\;cm$
\begin{equation}
\label{eq3}
a_{N} = \frac{{GM}}{{R^{2}}} = \frac{{\left( {6.67 \cdot 10^{ - 8}\;cm^{3}/g 
\cdot s^{2}} \right)\left( {2 \cdot 10^{33}\;g} \right)\;}}{{5.625 \cdot 
10^{29}\;cm^{2}}} = 2.37 \cdot 10^{ - 4}\;cm/s^{2}.
\end{equation}

Second, for the system (the Sun + spacecraft) we should determine smallness 
parameter $\lambda $, which is on the order of magnitude of the ratio 
between the characteristic velocity of the craft relative to the Sun and 
light velocity. For the above value of $R$:
\begin{equation}
\label{eq4}
\lambda \sim \sqrt{\frac{GM}{{c^{2}}R}} = \sqrt{\frac{{\left( {6.67 \cdot 
10^{ - 8}\;cm^{3}/g \cdot s^{2}} \right) \cdot \left( {2 \cdot 10^{33}\;g} 
\right)}}{{\left( {3 \cdot 10^{10}\;cm/s} \right)^{2} \cdot 1.5 \cdot 
10^{15}\;cm}}} \approx 10^{ - 5}.
\end{equation}

The above estimations show that the $a_{P} $ is not predictable by the 
general relativity either in the Newtonian or post-Newtonian (PN) 
approximation. If fact, the Einstein-Infeld-Hoffmann (EIH) corrections to 
acceleration differ from accelerations $a_{N} $ by a value of the order of 
$\sim \lambda ^{2}$, while relation $|a_{P} /a_{N} |$ is close to 
$\lambda^{1}$. The corrections to the acceleration in the PN approximation coincide 
with the distance and the $a_{P} $ is independent of distance.

It is our view that from the standpoint of explanation of the nature of 
$a_{P} $ the fact is essential that product $Hc$, where $H$ is Hubble 
constant, is equal to
\begin{equation}
\label{eq5}
Hc = 6.9 \cdot 10^{ - 8}\;cm/s^{2}.
\end{equation}
\noindent
which on the order of magnitude is close to $a_{P} $ (it is accepted that $H 
= 2.3 \cdot 10^{ - 18}\;1/s$), that is
\begin{equation}
\label{eq6}
a_{P} \approx - Hc.
\end{equation}
The closeness between $a_{P} $ and $Hc$ is noted both in ref. [1] and in 
many other papers. In this connection many people put forward the hypothesis 
that the additional acceleration $a_{P} $ owes its origin to the 
cosmological expansion of the Universe. However, no acceptable theoretical 
construction implementing this hypothesis has been proposed.

This paper makes an attempt to consider the motion of bodies in systems like 
(the Sun + spacecraft) within the framework of a so-called conformal 
geometrodynamics (CGD). CGD is a theoretical scheme based on equations that 
are a minimum conformally invariant generalization of Einstein equations for 
empty space. The CGD equations are derived in [3] (without $\lambda $ term) 
and in [4] (with $\lambda $ term). They are:
\begin{equation}
\label{eq7}
R_{\alpha \beta}  - \frac{1}{2} g_{\alpha \beta}  R = - 2A_{\alpha}  
A_{\beta}  - g_{\alpha \beta}  A^{2} - 2g_{\alpha \beta}  A^{\nu} _{;\nu}  + 
A_{\alpha ;\beta}  + A_{\beta ;\alpha}  + \lambda \cdot g_{\alpha \beta}  .
\end{equation}
Here $A_{\alpha}  ,\lambda $ are the vector and scalar fields that are 
attributes of the Riemannian variety dynamics. The form of equations (\ref{eq7}) is 
preserved in transformations
\begin{equation}
\label{eq8}
g_{\alpha \beta}  \to g_{\alpha \beta}  \cdot e^{2\sigma} ,\quad A_{\alpha}  
\to A_{\alpha}  - \sigma _{;\alpha}  ,\quad \lambda \to \lambda \cdot e^{ - 
2\sigma} ,
\end{equation}
\noindent
where $\sigma $ is an arbitrary scalar function of coordinates. These 
transformations pertain to the category of those, which effect the conformal 
mapping of Riemannian spaces and are termed conformal transformations (see, 
e.g., [5]).

To find solutions to equations (\ref{eq7}), coordinate conditions and gauge must be 
specified. For the coordinate conditions the well-known de Donder conditions 
will be used generalized so that they acquire the form invariant under 
transformations (\ref{eq8}). The conformally invariant form of the de Donder 
conditions is:
\begin{equation}
\label{eq9}
\frac{{g_{\alpha \lambda} } }{{\sqrt { - g}} }\left( {\sqrt { - g} 
\;g^{\lambda \sigma} } \right)_{,\sigma}  + 2A_{\sigma}  \frac{{g_{\alpha 
\lambda} } }{{\sqrt { - g}} }\left( {\sqrt { - g} \;g^{\lambda \sigma} } 
\right) = 0.
\end{equation}
For the gauge condition this paper uses condition
\begin{equation}
\label{eq10}
\lambda = \lambda _{0} = Const.
\end{equation}

When (\ref{eq10}) is valid, vector $A_{\alpha}  $ automatically satisfies Lorentz 
condition:
\begin{equation}
\label{eq11}
A^{\sigma}_{;\sigma}  = 0.
\end{equation}

  As will be demonstrated in this paper, the consideration of the body motion 
dynamics in a system like (the Sun + spacecraft) allows realization in a 
seminal manner of the hypothesis that the additional acceleration $a_{P} $ 
follows naturally from the CGD equations.

\subsection*{2. Equations and conditions in the EIH scheme used}

\bigskip

We will solve the problem of finding the equations of translational motion 
of two point bodies in the lower orders of approximation in parameter 
$\lambda \/ v/c$, where $v$ is the characteristic velocity of relative motion 
of the bodies. We will be solving with the Einstein-Infeld-Hoffmann (EIH) 
method in the form presented in refs. [8], [9]. When comparing specific 
expressions of this paper with the relevant expressions in refs. [8], [9] it 
should be kept in mind that this paper uses signature $\left( { - + + +}  
\right)$ and for the coordinate condition the de Donder condition in form 
(\ref{eq9}); the other notations are the same as those used in refs. [8], [9]. 

Assume that in our approximations $\lambda $ exerts no influence at all on 
the body motion and can be excluded from what follows\footnote{ For this 
reason hereafter there will be no danger to confuse function $\lambda \left( 
{x} \right)$with smallness parameter $\lambda $.} . 

In the EIH method, the arrangement of orders of smallness of $\gamma 
_{\alpha \beta}  $, $A_{\alpha}  $, where
\[
\gamma _{\alpha \beta}  \equiv h_{\alpha \beta}  - \frac{{1}}{{2}}\eta 
_{\alpha \beta}  \left( {\eta ^{\mu \nu} h_{\mu \nu} }  \right),
\quad
h_{\alpha \beta}  \equiv g_{\alpha \beta}  - \eta _{\alpha \beta}  ,
\]
$\eta _{\alpha \beta}  = {\tt diag}\left[ { - 1,1,1,1} \right]$ is the metric tensor 
of plane background space, depends on parameters of the bodies (masses, 
velocities) and distances between them. As applied to the system (the Sun + 
spacecraft) at distances (\ref{eq1}), this consideration adopts the following order 
of smallness arrangement:
\begin{equation}
\label{eq12}
\left. {\begin{array}{l}
 {\gamma _{00} = \mathop {\gamma} \limits_{2} {}_{00} + \mathop {\gamma 
}\limits_{4} {}_{00} + \mathop {\gamma} \limits_{5} {}_{00} + ...,} \\ 
 {\gamma _{0k} = \mathop {\gamma} \limits_{3} {}_{0k} + \mathop {\gamma 
}\limits_{4} {}_{0k} + \mathop {\gamma} \limits_{5} {}_{0k} + ...,} \\ 
 {\gamma _{mn} = \mathop {\gamma} \limits_{4} {}_{mn} + \mathop {\gamma 
}\limits_{5} {}_{mn} + \mathop {\gamma} \limits_{6} {}_{mn} + ...\;,} \\ 
 {A_{0} = \mathop {A}\limits_{3} {}_{0} + \mathop {A}\limits_{4} {}_{0} + 
...,} \\ 
 {A_{k} = \mathop {A}\limits_{4} {}_{k} + \mathop {A}\limits_{5} {}_{k} + 
...\;.} \\ 
 \end{array}}  \right\}
\end{equation}
As it will follow from the following, the above order of smallness 
arrangement rules out the influence of field $A_{\alpha}  $ on spacecraft 
motion in the Newtonian approximation. 

Equations (\ref{eq7}) lead to the following equations for $\gamma _{\alpha \beta}  $ 
(only those equations are written out which are used in what follows):
\begin{equation}
\label{eq13}
\left[ {00;\lambda^{2}} \right] \Rightarrow \quad \quad - 
\frac{{1}}{{2}}\left( {\Delta \mathop {\gamma} \limits_{2} {}_{00}}  \right) 
= 0.
\end{equation}
\begin{equation}
\label{eq14}
\left[ {0k;\lambda ^{3}} \right] \Rightarrow \quad \quad - 
\frac{{1}}{{2}}\left( {\Delta \mathop {\gamma} \limits_{3} {}_{0k}}  \right) 
+ \frac{{1}}{{2}}\left( { - \mathop {\gamma} \limits_{2} {}_{00,} {}\mathop 
{_{0}} \limits_{1} + \mathop {\gamma} \limits_{3} {}_{0s,s}}  \right)_{,k} = 
0.
\end{equation}
\begin{equation}
\label{eq15}
\left\{ {\begin{array}{l}
 {\left[ {mn;\lambda^{4}} \right] \Rightarrow}  \\ 
 {\frac{{1}}{{2}}\left\{ {\left( { - \mathop {\gamma} \limits_{3} {}_{0m,} 
{}\mathop {_{0}} \limits_{1} {}_{n} + \mathop {\gamma} \limits_{4} {}_{ms,sn}}  
\right)} \right. + \left( { - \mathop {\gamma} \limits_{3} {}_{0n,} {}\mathop 
{_{0}} \limits_{1} {}_{m} + \mathop {\gamma} \limits_{4} {}_{ns,sm}}  \right) 
- \Delta \mathop {\gamma} \limits_{4} {}_{mn} -}  \\ 
 { - \left. {\delta _{mn} \left( {\mathop {\gamma} \limits_{2} {}_{00,} 
{}\mathop {_{0}} \limits_{1} {}\mathop {_{0}} \limits_{1} - 2\mathop {\gamma 
}\limits_{3} {}_{0s,} {}\mathop {_{0}} \limits_{1} {}_{s} + \mathop {\gamma 
}\limits_{4} {}_{pq,pq}}  \right)} \right\} + \frac{{1}}{{4}}\mathop {\gamma 
}\limits_{2} {}_{00} \mathop {\gamma} \limits_{2} {}_{00,mn} +}  \\ 
 { + \frac{{1}}{{8}}\mathop {\gamma} \limits_{2} {}_{00,m} \mathop {\gamma 
}\limits_{2} {}_{00,n} - \frac{{1}}{{8}}\delta _{mn} \mathop {\gamma 
}\limits_{2} {}_{00} \left( {\Delta \mathop {\gamma} \limits_{2} {}_{00}}  
\right) - \frac{{3}}{{16}}\delta _{mn} \mathop {\gamma} \limits_{2} {}_{00,s} 
\mathop {\gamma} \limits_{2} {}_{00,s} = 0.} \\ 
 \end{array}}  \right.
\end{equation}
\begin{equation}
\label{eq16}
\left[ {0k;\lambda^{4}} \right] \Rightarrow \quad \quad - 
\frac{{1}}{{2}}\left( {\Delta \mathop {\gamma} \limits_{4} \;_{0k}}  \right) 
+ \frac{{1}}{{2}}\left( { - \mathop {\gamma} \limits_{3} {}_{00,} {}\mathop 
{_{0}} \limits_{1} + \mathop {\gamma} \limits_{4} \;_{0s,s}}  \right)_{,k} = 
\mathop {A}\limits_{4} {}_{0,k} .
\end{equation}
\begin{equation}
\label{eq17}
\left[ {00;\lambda^{4}} \right] \Rightarrow \quad \quad - 
\frac{{1}}{{2}}\Delta \mathop {\gamma} \limits_{4} {}_{00} + 
\frac{{1}}{{2}}\mathop {\gamma} \limits_{4} {}_{pq,pq} + 
\frac{{3}}{{16}}\mathop {\gamma} \limits_{2} {}_{00,s} \mathop {\gamma 
}\limits_{2} {}_{00,s} + \frac{{3}}{{8}}\mathop {\gamma} \limits_{2} {}_{00} 
\left( {\Delta \mathop {\gamma} \limits_{2} {}_{00}}  \right) = 0.
\end{equation}
\begin{equation}
\label{eq18}
\left\{ {\begin{array}{l}
 {\left[ {mn;\lambda ^{5}} \right] \Rightarrow}  \\ 
 {\frac{{1}}{{2}}\left\{ {\left( { - \mathop {\gamma} \limits_{4} {}_{0m,} 
{}\mathop {_{0}} \limits_{1} {}_{n} + \mathop {\gamma} \limits_{5} {}_{ms,sn}}  
\right)} \right. + \left( { - \mathop {\gamma} \limits_{4} {}_{0n,} {}\mathop 
{_{0}} \limits_{1} {}_{m} + \mathop {\gamma} \limits_{5} {}_{ns,sm}}  \right) 
- \Delta \mathop {\gamma}\limits_{5} {}_{mn}} - \\
- \left. {\delta _{mn} \left( { 
- 2\mathop {\gamma} \limits_{4} {}_{0s,} {}\mathop {_{0}} \limits_{1} {}_{s} + 
\mathop {\gamma} \limits_{5} {}_{pq,pq}}  \right)} \right\} \\ 
 { = \mathop {A}\limits_{5} {}_{m,n} + \mathop {A}\limits_{5} {}_{n,m} - 
2\delta _{mn} \left[ { - \mathop {A}\limits_{4} {}_{0,0} + \mathop 
{A}\limits_{5} {}_{l,l}}  \right].} \\ 
 \end{array}}  \right.
\end{equation}

\noindent
Coordinate conditions (\ref{eq9}) are of form (in square brackets: $C.C. = 
$Coordinate Condition):
\begin{equation}
\label{eq19}
\left[ {C.C.;0;\lambda ^{3}} \right] \Rightarrow \quad \quad \mathop {\gamma 
}\limits_{3} {}_{0l,l} = \mathop {\gamma} \limits_{2} {}_{00,} \mathop {_{0} 
}\limits_{1} + 2\mathop {A}\limits_{3} {}_{0} .
\end{equation}
\begin{equation}
\label{eq20}
\left[ {C.C.;k;\lambda ^{4}} \right] \Rightarrow \quad \quad \mathop {\gamma 
}\limits_{4} {}_{kl,l} = \mathop {\gamma} \limits_{3} {}_{0k,} \mathop {_{0} 
}\limits_{1} - \frac{{1}}{{4}}\mathop {\gamma} \limits_{2} {}_{00} \mathop 
{\gamma} \limits_{2} {}_{00,k} .
\end{equation}
\begin{equation}
\label{eq21}
\left[ {C.C.;0;\lambda ^{4}} \right] \Rightarrow \quad \quad \mathop {\gamma 
}\limits_{4} {}_{0l,l} = 2\mathop {A}\limits_{4} {}_{0} .
\end{equation}
\begin{equation}
\label{eq22}
\left[ {C.C.;0;\lambda^{5}} \right] \Rightarrow \quad \mathop {\gamma 
}\limits_{5} {}_{0l,l} = \mathop {\gamma} \limits_{4} {}_{00,} \mathop {_{0} 
}\limits_{1} - \frac{{1}}{{2}}\mathop {\gamma} \limits_{3} {}_{0p} \mathop 
{\gamma} \limits_{2} {}_{00,p} - \frac{{1}}{{4}}\mathop {\gamma} \limits_{2} 
{}_{00} \mathop {\gamma} \limits_{2} {}_{00,0} + \frac{{1}}{{2}}\mathop 
{\gamma} \limits_{2} {}_{00} \mathop {\gamma} \limits_{3} {}_{0s,s} + 2\mathop 
{A}\limits_{5} {}_{0} .
\end{equation}
\begin{equation}
\label{eq23}
\left[ {C.C.;k;\lambda^{5}} \right] \Rightarrow \quad \quad \mathop {\gamma 
}\limits_{4} {}_{0k,} \mathop {_{0}} \limits_{1} - \mathop {\gamma 
}\limits_{5} {}_{kl,l} + 2\mathop {A}\limits_{5} {}_{k} = 0.
\end{equation}

Among all conditions (\ref{eq11}), only the gauge conditions of the lower 
approximations are sufficient (in square brackets: $G.C. = $Gauge 
Condition):
\begin{equation}
\label{eq24}
\left[ {G.C.;\lambda ^{4}} \right] \Rightarrow \quad \quad \mathop 
{A}\limits_{4} {}_{l,l} = 0.
\end{equation}
\begin{equation}
\label{eq25}
\left[ {G.C.;\lambda^{5}} \right] \Rightarrow \quad \quad - \mathop 
{A}\limits_{4} {}_{0,0} + \mathop {A}\limits_{5} {}_{l,l} = 0.
\end{equation}

From equations (\ref{eq7}) it follows that functions $\mathop {A}\limits_{3} {}_{0} 
{}\mathop {A}\limits_{4} {}_{0} ,\;\mathop {A}\limits_{4} {}_{k} ,\;\mathop 
{A}\limits_{5} {}_{k} $ should satisfy the following equations:
\begin{equation}
\label{eq26}
\left[ {0;\lambda ^{3}} \right] \Rightarrow \quad \quad - \Delta \mathop 
{A}\limits_{3} {}_{0} = 0.
\end{equation}
\begin{equation}
\label{eq27}
\left[ {0;\lambda ^{4}} \right] \Rightarrow \quad \quad - \Delta \mathop 
{A}\limits_{4} {}_{0} = 0.
\end{equation}
\begin{equation}
\label{eq28}
\left[ {k;\lambda ^{4}} \right] \Rightarrow \quad \quad - \Delta \mathop 
{A}\limits_{4} {}_{k} = 0.
\end{equation}
\begin{equation}
\label{eq29}
\left[ {k;\lambda ^{5}} \right] \Rightarrow \quad \quad - \Delta \mathop 
{A}\limits_{5} {}_{k} = 0.
\end{equation}

Thus, functions $\mathop {A}\limits_{3} {}_{0} ,\;\mathop {A}\limits_{4} 
{}_{0} ,\;\mathop {A}\limits_{4} {}_{k} ,\;\mathop {A}\limits_{5} {}_{k} $ 
should be harmonic functions.

\subsection*{3. Centrally symmetric static solution to CGD equations}

\bigskip

The centrally symmetric static (CSS) solution to the CGD equations is 
presented in different forms in several papers (see [6], [7]). Here the CSS 
solution will be presented in a form convenient for comparison to the 
Schwarzschild solution.

In the CSS problem the squared interval is
\[
ds^{2} = - {\tt exp}\left( {\gamma}  \right) \cdot dt^{2} + 
{\tt exp}\left( {\alpha}  
\right) \cdot dz^{2} + {\tt exp}\left( {\beta}  \right) \cdot \left( {d\theta ^{2} 
+ {\tt sin}^{2}\theta \cdot d\varphi^{2}} \right).
\]
In this case the gauge vector can have as few as two nonzero, radial 
coordinate dependent components:
\[
A_{\alpha}  = \left( {\phi ,f,0,0} \right).
\]
Having made use of the gauge transformation with radial variable dependent 
function $\sigma = \sigma \left( {z} \right)$, set $A_{1} = f$ equal to 
zero. Then replace the radial coordinate so that condition $g_{00} g_{11} = 
- 1$ be met. This transformation does not result in appearance of $A_{1} 
$-component of field $A_{\alpha}  $. Upon the above two transformations, 
without loss of generality, the metric can be written in the form
\[
ds^{2} = - {\tt exp}\left( {\gamma}  \right) \cdot dt^{2} + {\tt exp}\left( { - \gamma}  
\right) \cdot dz^{2} + {\tt exp}\left( {\beta}  \right) \cdot \left( {d\theta ^{2} 
+ {\tt sin}^{2}\theta \cdot d\varphi ^{2}} \right)
\]
\noindent
and vector $A_{\alpha}  $ in the form 
\[
A_{\alpha}  = \left( {\phi ,0,0,0} \right).
\]

Four equations are obtained for four functions $\gamma ,\;\beta ,\;\phi 
,\;\lambda $. 

Equation $G_{0}^{0} = T_{0}^{0} \Rightarrow $
\[
{\tt exp}\left( {\gamma}  \right) \cdot \left[ {\beta '' + {\textstyle{{3} \over 
{4}}}\left( {\beta '} \right)^{2} + {\textstyle{{1} \over {2}}}\gamma '\beta 
'} \right] - {\tt exp}\left( { - \beta}  \right) = 3{\tt exp}\left( { - \gamma}  \right) 
\cdot \phi ^{2} + \lambda .
\]
Equation $G_{1}^{1} = T_{1}^{1} \Rightarrow $
\[
{\tt exp}\left( {\gamma}  \right) \cdot \left[ {{\textstyle{{1} \over {4}}}\left( 
{\beta '} \right)^{2} + {\textstyle{{1} \over {2}}}\gamma '\beta '} \right] 
- {\tt exp}\left( { - \beta}  \right) = {\tt exp}\left( { - \gamma}  \right) \cdot \phi 
^{2} + \lambda .
\]
Equation $G_{2}^{2} = T_{2}^{2} \Rightarrow $
\[
{\tt exp}\left( {\gamma}  \right) \cdot \left[ {{\textstyle{{1} \over {2}}}\beta 
'' + {\textstyle{{1} \over {4}}}\left( {\beta '} \right)^{2} + 
{\textstyle{{1} \over {2}}}\gamma '' + {\textstyle{{1} \over {2}}}\left( 
{\gamma '} \right)^{2} + {\textstyle{{1} \over {2}}}\beta '\gamma '} \right] 
= {\tt exp}\left( { - \gamma}  \right) \cdot \phi ^{2} + \lambda .
\]
Equation $G_{0}^{1} = T_{0}^{1} \Rightarrow $
\[
0 = \phi ' - \gamma '\phi .
\]

It turns out that there are three solution types. We are concerned about 
that type, which contains the Schwarzschild solution as a special case. Here 
is the solution type with the procedure itself of its finding being omitted. 
\begin{equation}
\label{eq30}
\left. {\begin{array}{l}
 {\phi = p_{0} \cdot {\tt exp}\left( {\gamma}  \right)} \\ 
 {{\tt exp}\left( {\beta}  \right) = A_{0} \cdot {\tt sh}^{2}\left( {p_{0} z + a_{0}}  
\right)} \\ 
 {{\tt exp}\left( {\gamma}  \right) = \frac{{1}}{{p_{0}^{2} A_{0}} } + B_{0} \cdot 
\left[ {p_{0} z \cdot {\tt cth}\left( {p_{0} z + a_{0}}  \right) - 1} \right] + 
b_{0} p_{0} \cdot {\tt cth}\left( {p_{0} z + a_{0}}  \right)} \\ 
 {\lambda \left( {z} \right) = B_{0} p_{0}^{2}}  \\ 
 \end{array}}  \right\}
\end{equation}
\noindent
The zero-subscripted quantities are integration constants.

With appropriately chosen constants a solution of type (\ref{eq30}) can approximate 
the Schwarzschild solution as closely as is wished in a certain range of the 
radial variable. To obtain the approximation, set the following in (\ref{eq30}):
\begin{equation}
\label{eq31}
B_{0} = 0,\quad a_{0} = 0,\quad A_{0} p_{0}^{2} = 1.
\end{equation}
Then for $e^{\gamma} $:
\begin{equation}
\label{eq32}
e^{\gamma}  = 1 + b_{0} p_{0} \cdot {\tt cth}\left( {p_{0} z} \right).
\end{equation}
Having introduced notation $b_{0} \equiv - r_{0} $, $p_{0} \equiv 1/L$, 
write the CSS solution. 
\begin{equation}
\label{eq33}
e^{\gamma}  = 1 - \frac{{r_{0}} }{{L}} \cdot {\tt cth}\left( {\frac{{z}}{{L}}} 
\right),
\quad
e^{\beta}  = L^{2} \cdot {\tt sh}^{2}\left( {\frac{{z}}{{L}}} \right),
\end{equation}
\begin{equation}
\label{eq34}
\phi = \frac{{1}}{{L}}\left[ {1 - \frac{{r_{0}} }{{L}} \cdot {\tt cth}\left( 
{\frac{{z}}{{L}}} \right)} \right].
\end{equation}

In the range of the radial variable
\begin{equation}
\label{eq35}
\left( {r_{0} /L} \right) < \left( {z/L} \right) \ll 1
\end{equation}

\noindent
expressions (\ref{eq33}), (\ref{eq34}) assume the following form:
\begin{equation}
\label{eq36}
e^{\gamma}  = 1 - \frac{{r_{0}} }{{z}},
\quad
e^{\beta}  = z^{2},
\end{equation}
\begin{equation}
\label{eq37}
\phi = \frac{{1}}{{L}}\left[ {1 - \frac{{r_{0}} }{{z}}} \right].
\end{equation}

Expressions (\ref{eq36}) are the same as the associated expressions in the 
Schwarz\-schild solution. In so doing $r_{0} $ has the meaning of 
gravitational radius. As for (\ref{eq37}), there is no analog of this expression in 
the Schwarzschild solution. In expressions (\ref{eq36}), (\ref{eq37}), two facts seem 
essential. First, in the range of radial variable (\ref{eq35}), for its description 
the CGD equation solution requires not only dimensional constant $r_{0} $, 
but also one more dimensional constant $L$. Second, in the above-mentioned 
range of radial variable the principal term in the expansion of function 
$\phi $ is constant $1/L$.

\subsection*{4. Finding functions $\gamma _{\alpha \beta}  $, $A_{\alpha}  $ with the EIH 
method}

\bigskip

For us it is convenient first to find the equations of translational motion 
of two particles under the assumption that the particle masses are 
commensurable in magnitude, that is, to obtain results similar to those 
obtained in refs. [8], [9], however not for the equations of general 
relativity, but for equations (\ref{eq7}). Then from the found equations of motion 
the equations of motion can be obtained straightforwardly for the system, in 
which one of the particles is a trial particle. 

The general approach to the solution of our problem is that components 
$\gamma _{\alpha \beta}  $ will be represented in form $\gamma _{\alpha 
\beta}  = \bar {\gamma} _{\alpha \beta}  + \delta \gamma _{\alpha \beta}  $. 
Here $\bar {\gamma} _{\alpha \beta}  $ are expressions that are obtained in 
the framework of general relativity for two point particles of masses $M,m$ 
without inclusion of additional terms of $A_{\alpha}  $. On the other hand, 
$\delta \gamma _{\alpha \beta}  $ is an addition determined entirely by the 
$A_{\alpha}  $.

The structure of all quantities $\gamma _{\alpha \beta}  $, $A_{\alpha}  $ 
is essentially predetermined in the scheme under discussion by expressions 
for $\gamma _{00} $ and $A_{0} $ in the lower orders of smallness. In our 
case the expressions for $\mathop {\gamma} \limits_{2} {}_{00} $, $\mathop 
{\gamma} \limits_{3} {}_{00} $ should coincide with expressions
\begin{equation}
\label{eq38}
\mathop {\gamma} \limits_{2} {}_{00} = 4\frac{{M}}{{r_{1}} } + 
4\frac{{m}}{{r_{2}} },
\quad
\mathop {\gamma} \limits_{3} {}_{00} = 0,
\end{equation}

\noindent
that is with those expressions, which in the case of general relativity lead 
to the well-known equations of translational motion in the PN approximation.

As for $A_{0} $, certain considerations regarding its choice follow from 
approximate expressions (\ref{eq36}), (\ref{eq37}) for the exact CSS solution to the CGD 
equations. Clear that the expansion of $A_{0} $ should begin with the 
constant $1/L$, where $L$ is the dimensional parameter used to determine 
asymptotic $A_{0} $ at long distances. When considering multi-particle 
problems, the expansion of the exact one-particle solution does not allow 
detection of the presence of particle velocity containing terms. Appearance 
of this type of terms cannot also be excluded in the expression for $A_{0} 
$, when a system of at least two particles is considered. Note that a 
similar situation arises, for example, when one attempts to find the Kerr 
solution with the EIH method [9] or tries to determine radiation friction 
force in the Dirac-Lorentz equation with this method [10]. 

Reasoning from the above considerations, write the expression for $A_{0} $, 
which could be composed of the quantities present in the problem. The 
desired expression:

- should begin with constant $1/L$, 

- can contain particle velocities $\dot {\xi} _{k} ,\;\dot {\eta} _{k} $ in 
the next order of smallness, 

- should be, according to (\ref{eq26}), (\ref{eq27}), a harmonic function, 

- if it contains a part depending on coordinates, then the part should be 
decreasing with distance from the system, 

- should be symmetric about replacement of parameters of one particle by 
those of the other. 

It turns out that all of these requirements are satisfied, if $A_{0} $ is 
represented as
\begin{equation}
\label{eq39}
A_{0} = \mathop {A}\limits_{3} {}_{0} + \mathop {A}\limits_{4} {}_{0} = 
\frac{{1}}{{L}} + \kappa \cdot \frac{{\mu R^{2}}}{{\left( {m + M} 
\right)L}}\left\{ {\frac{{\left( {X_{l} \dot {\xi} _{l}}  
\right)}}{{r_{1}^{3}} } + \frac{{\left( {\left( {x_{l} - \eta _{l}}  
\right)\dot {\eta} _{l}}  \right)}}{{r_{2}^{3}} }} \right\}
\end{equation}

\noindent
and $R/L$ is assumed to be of the third order of smallness. Here $\kappa $ 
is a constant numerical coefficient, $\mu = mM/\left( {m + M} \right)$ is 
scaled mass of a system of two particles. In this consideration multiplier 
$\mu /\left( {m + M} \right)$ is assumed to be of the order of unity. The 
need for the introduction of $\kappa $ to the $A_{0} $ construction is due 
to the lack of uniqueness that is inherent in the EIH method in construction 
of approximations to the exact solution with using it (for more details, see 
[9], [10]). The considerations on selection of the value of the constant 
$\kappa $ will be presented later. 

 $A_{k} $ is found from gauge conditions (\ref{eq24}), (\ref{eq25}). From equation (\ref{eq28}) and 
condition (\ref{eq24}) it follows that $\mathop {A}\limits_{4} {}_{k} = 0$. Write the 
expression for $A_{k} $ in the vicinity of the first particle alone; in so 
doing make use of the equation of motion of particle in the Newtonian 
approximation.
\begin{equation}
\label{eq40}
\begin{array}{l}
A_{k} = \mathop {A}\limits_{4} {}_{k} + \mathop {A}\limits_{5} {}_{k} = \\
= \kappa 
\cdot \frac{{\mu} }{{\left( {m + M} \right)L}}\left\{ { - \frac{{mR_{k} 
}}{{Rr_{1}} } - 2\frac{{\left( {R_{l} \dot {R}_{l}}  \right)\dot {\xi} _{k} 
}}{{r_{1}} } - \frac{{R^{2}\left( {X_{l} \dot {\xi} _{l}}  \right)\dot {\xi 
}_{k}} }{{r_{1}^{3}} }} \right\} + \left( {1 
\mathbin{\lower.3ex\hbox{$\buildrel\textstyle\rightarrow\over 
{\smash{\leftarrow}\vphantom{_{\vbox to.5ex{\vss}}}}$}} 2}\right) \\
\end{array} 
\end{equation}
From here on the term with two arrows $\left( {1 
\mathbin{\lower.3ex\hbox{$\buildrel\textstyle\rightarrow\over 
{\smash{\leftarrow}\vphantom{_{\vbox to.5ex{\vss}}}}$}} 2} \right)$ denotes 
the expression derived from the written one through replacement of 
parameters of one particle by those of the other.

The substitution of expression (\ref{eq39}) for $A_{0} $ into equations (\ref{eq14}), (\ref{eq16}) 
with taking into account (\ref{eq19}), (\ref{eq21}) yields the following form of the 
expression for $\delta \gamma {}_{0k} $:
\begin{equation}
\label{eq41}
\delta \gamma {}_{0k} = \mathop {\delta \gamma} \limits_{3} {}_{0k} + \mathop 
{\delta \gamma} \limits_{4} {}_{0k} = \frac{{2X_{k}} }{{3L}} - \kappa 
\frac{{\mu R^{2}}}{{\left( {m + M} \right)L}} \cdot \frac{{\dot {\xi} _{k} 
}}{{r_{1}} } + \left( {1 
\mathbin{\lower.3ex\hbox{$\buildrel\textstyle\rightarrow\over 
{\smash{\leftarrow}\vphantom{_{\vbox to.5ex{\vss}}}}$}} 2} \right).
\end{equation}

Found expressions (\ref{eq39}), (\ref{eq40}), (\ref{eq41}) are sufficient to determine the form of 
the correction to the Newtonian equation of particle motion. These 
corrections are obtained by integration of equation (\ref{eq18}) over 
infinitesimal-radius spheres surrounding the point particles. Thus we obtain 
the correction to the equations of motion for the first particle.

Write equation (\ref{eq18}) with having rearranged all the terms to the left side.
\begin{equation}
\label{eq42}
\begin{array}{l}
 \frac{{1}}{{2}}\left\{ {\left( { - \mathop {\gamma} \limits_{4} {}_{0m,} 
{}\mathop {_{0}} \limits_{1} {}_{n} + \mathop {\gamma} \limits_{5} {}_{ms,sn}}  
\right)} \right. + \left( { - \mathop {\gamma} \limits_{4} {}_{0n,} {}\mathop 
{_{0}} \limits_{1} {}_{m} + \mathop {\gamma} \limits_{5} {}_{ns,sm}}  \right) 
-\\
- \Delta \mathop {\gamma} \limits_{5} {}_{mn} - \left. {\delta _{mn} \left( { 
- 2\mathop {\gamma} \limits_{4} {}_{0s,} {}\mathop {_{0}} \limits_{1} {}_{s} + 
\mathop {\gamma} \limits_{5} {}_{pq,pq}}  \right)} \right\} \\ 
 - \mathop {A}\limits_{5} {}_{m,n} - \mathop {A}\limits_{5} {}_{n,m} + 2\delta 
_{mn} \left[ { - \mathop {A}\limits_{4} {}_{0,0} + \mathop {A}\limits_{5} 
{}_{l,l}}  \right] = 0. \\ 
 \end{array}
\end{equation}

In the integration of this equation the total contribution from the terms 
containing second derivatives of components $\mathop {\delta \gamma 
}\limits_{5} {}_{mn} $ is zero, as they form curl combinations. The 
contribution from the following combination is zero for the same reason:
\[
\frac{{1}}{{2}}\left\{ {\left( {\delta _{mn} \mathop {\gamma} \limits_{4} 
{}_{0s,} {}\mathop {_{0}} \limits_{1} {}_{s} - \mathop {\gamma} \limits_{4} 
{}_{0n,} {}\mathop {_{0}} \limits_{1} {}_{m}}  \right)} \right\}.
\]
The contribution from 
\[
 + 2\delta _{mn} \left( { - \mathop {A}\limits_{4} {}_{0,0} + \mathop 
{A}\limits_{5} {}_{l,l}}  \right)
\]

\noindent
is zero by virtue of the gauge condition. As a result, to find the 
correction to the equation of motion, the contribution from the following 
terms must be calculated:
\[
 - \frac{{1}}{{2}}\mathop {\gamma} \limits_{4} {}_{0m,} {}\mathop {_{0} 
}\limits_{1} {}_{n} + \frac{{1}}{{2}}\delta _{mn} \mathop {\gamma 
}\limits_{4} {}_{0s,} {}\mathop {_{0}} \limits_{1} {}_{s} - \mathop 
{A}\limits_{5} {}_{m,n} - \mathop {A}\limits_{5} {}_{n,m} .
\]

This correction will therewith enter into the equation of motion 
\begin{equation}
\label{eq43}
2M\ddot {\xi} _{k} = \frac{{mM}}{{R^{3}}}R_{k} + \Im _{k} 
\end{equation}

\noindent
as addition $\Im _{k} $,

\begin{equation}
\label{eq44}
\Im _{k} = \frac{{1}}{{4\pi} }\oint {\left\{ { - \frac{{1}}{{2}}\mathop 
{\gamma} \limits_{4} {}_{0k,} {}\mathop {_{0}} \limits_{1} {}_{n} + 
\frac{{1}}{{2}}\delta _{kn} \mathop {\gamma} \limits_{4} {}_{0s,} {}\mathop 
{_{0}} \limits_{1} {}_{s} - \mathop {A}\limits_{5} {}_{k,n} - \mathop 
{A}\limits_{5} {}_{n,k}}  \right\}ds_{n}}  .
\end{equation}

The contributions of separate terms to surface integral (\ref{eq44}) are:
\begin{equation}
\label{eq45}
\frac{{1}}{{4\pi} }\oint {\left\{ { - \frac{{1}}{{2}}\mathop {\gamma 
}\limits_{4} {}_{0k,} {}\mathop {_{0}} \limits_{1} {}_{n}}  \right\}ds_{n}}  = 
- \kappa \cdot \frac{{\mu} }{{2\left( {m + M} \right)L}} \cdot \left( 
{\frac{{mR_{k}} }{{R}} + 2\left( {R_{l} \dot {R}_{l}}  \right)\dot {\xi 
}_{k}}  \right).
\end{equation}
\begin{equation}
\label{eq46}
\begin{array}{l}
 \frac{{1}}{{4\pi} }\oint {\left\{ {\frac{{1}}{{2}}\delta _{kn} \mathop 
{\gamma} \limits_{4} {}_{0s,} {}\mathop {_{0}} \limits_{1} {}_{s}}  
\right\}ds_{n}}  = \frac{{1}}{{4\pi} }\oint {\left\{ {\delta _{kn} \mathop 
{A}\limits_{4} {}_{0,0}}  \right\}ds_{n}}  \\ 
 = \kappa \cdot \frac{{\mu} }{{2\left( {m + M} \right)L}} \cdot \left( 
{\frac{{mR_{k}} }{{3R}} + \frac{{2}}{{3}}\left( {R_{l} \dot {R}_{l}}  
\right)\dot {\xi} _{k}}  \right). \\ 
 \end{array}
\end{equation}
\begin{equation}
\label{eq47}
\begin{array}{l}
 \frac{{1}}{{4\pi} }\oint {\left\{ { - \mathop {A}\limits_{5} {}_{k,n} - 
\mathop {A}\limits_{5} {}_{n,k}}  \right\}ds_{n}}  = \kappa \cdot \left( { - 
\frac{{4}}{{3}} \cdot \frac{{M}}{{L}} \cdot \frac{{R_{k}} }{{R}} - 
\frac{{8}}{{3}} \cdot \frac{{M}}{{mL}} \cdot \left( {R_{l} \dot {R}_{l}}  
\right)\dot {\xi} _{k}}  \right) \\ 
 = - \kappa \cdot \frac{{\mu} }{{\left( {m + M} \right)L}} \cdot \left( { - 
\frac{{4}}{{3}} \cdot \frac{{mR_{k}} }{{R}} + \frac{{8}}{{3}}\left( {R_{l} 
\dot {R}_{l}}  \right)\dot {\xi} _{k}}  \right). \\ 
 \end{array}.
\end{equation}

Substitution of (\ref{eq45})-(\ref{eq47}) into (\ref{eq44}) gives:
\begin{equation}
\label{eq48}
\Im _{k} = \kappa \cdot \frac{{\mu} }{{\left( {m + M} \right)L}} \cdot 
\frac{{mR_{k}} }{{R}} + 2\kappa \cdot \frac{{\mu} }{{\left( {m + M} 
\right)L}} \cdot \left( {R_{l} \dot {R}_{l}}  \right)\dot {\xi} _{k} .
\end{equation}

As a result, from (\ref{eq43}), (\ref{eq44}), (\ref{eq48}) we obtain the following correction to 
acceleration $\delta \ddot {\xi} _{k} $:
\begin{equation}
\label{eq49}
\delta \ddot {\xi} _{k} = \kappa \cdot \frac{{\mu m}}{{2\left( {m + M} 
\right)M}} \cdot \frac{{R_{k}} }{{LR}} + \kappa \cdot \frac{{\mu} }{{\left( 
{m + M} \right)ML}} \cdot \left( {R_{l} \dot {R}_{l}}  \right)\dot {\xi 
}_{k} .
\end{equation}

In terms of derivatives with respect to regular time and masses in regular 
units of measurement equation (\ref{eq49}) is written as:
\begin{equation}
\label{eq50}
\frac{{d^{2}\delta \xi _{k}} }{{dt^{2}}} = \kappa \cdot \frac{{\mu 
m}}{{2\left( {m + M} \right)M}} \cdot \frac{{c^{2}R_{k}} }{{LR}} + \kappa 
\cdot \frac{{\mu c^{2}}}{{\left( {m + M} \right)MLG}} \cdot \left( {R_{l} 
\frac{{dR_{l}} }{{dt}}} \right)\frac{{d\xi _{k}} }{{dt}}.
\end{equation}

Assume that constant $L$ is of cosmological origin and is related with 
Hubble constant $H$ as
\begin{equation}
\label{eq51}
L = c/H.
\end{equation}

Having substituted (\ref{eq51}) into (\ref{eq50}), we obtain:
\begin{equation}
\label{eq52}
\frac{{d^{2}\delta \xi _{k}} }{{dt^{2}}} = \frac{{\kappa \mu m}}{{2\left( {m 
+ M} \right)M}} \cdot cH\frac{{R_{k}} }{{R}} + \frac{{\kappa \mu} }{{\left( 
{m + M} \right)MG}} \cdot cH \cdot \left( {R_{l} \frac{{dR_{l}} }{{dt}}} 
\right)\frac{{d\xi _{k}} }{{dt}}.
\end{equation}

Expression (\ref{eq52}) is just the correction to the Newtonian expression for 
acceleration which follows from the CGD equations under the assumptions 
specified at the beginning of this section.

\subsection*{5. Equations of motion of spacecrafts Pioneer 10, 11}

\bigskip

Write equations (\ref{eq52}) in the special case, where mass of the first particle 
is much less than that of the second particle, that is where the first 
particle can be considered as a trial particle. In this case
\begin{equation}
\label{eq53}
\frac{{d^{2}\delta \xi _{k}} }{{dt^{2}}} = \frac{{\kappa} }{{2}} \cdot 
cH\frac{{R_{k}} }{{R}} - \frac{{\kappa} }{{mG}} \cdot cH \cdot \left( {R_{l} 
\frac{{d\xi _{l}} }{{dt}}} \right)\frac{{d\xi _{k}} }{{dt}}.
\end{equation}

It is interesting to note that the first term in the right-hand side of (\ref{eq53}) 
depends neither on masses of the bodies nor on the distance between them 
whatsoever. 

From (\ref{eq53}) it follows that in the trial particle motion along circumference 
the correction from the second term in the right-hand side of (\ref{eq53}) is zero, 
the correction to the Newtonian expression for the acceleration is 
determined by the first term.
\begin{equation}
\label{eq54}
\frac{{d^{2}\delta \xi _{k}} }{{dt^{2}}} = \frac{{\kappa} }{{2}} \cdot 
cH\frac{{R_{k}} }{{R}}.
\end{equation}

Consider motion of the trial particle along radius. We will omit subscript 
``1'' in writing the radial vector values. In this case we obtain for the 
radial component of acceleration from (\ref{eq53}):
\begin{equation}
\label{eq55}
\frac{{d^{2}\delta \xi} }{{dt^{2}}} = \frac{{\kappa} }{{2}} \cdot cH - 
\frac{{\kappa} }{{mG}} \cdot cH \cdot R\left( {\frac{{d\xi 
}}{{dt}}\frac{{d\xi} }{{dt}}} \right).
\end{equation}

Write the squared radial velocity with making use of the law of conservation 
of energy, that is as
\[
\left( {\frac{{d\xi} }{{dt}}\frac{{d\xi} }{{dt}}} \right) = \frac{{2E_{0} 
}}{{M}} + \frac{{2Gm}}{{R}}.
\]

We arrive at:
\begin{equation}
\label{eq56}
\frac{{d^{2}\delta \xi} }{{dt^{2}}} = \frac{{\kappa} }{{2}} \cdot cH - 
\frac{{\kappa} }{{mG}} \cdot cH \cdot R \cdot \left( {\frac{{2E_{0}} }{{M}} 
+ \frac{{2Gm}}{{R}}} \right) = - \frac{{3\kappa} }{{2}} \cdot cH - 
\frac{{2\kappa RE_{0}} }{{mMG}} \cdot cH.
\end{equation}

Consider the spacecrafts as trial particles. Total energy $E_{0} $ of the 
spacecraft leaving the system with minimum kinetic energy is close to zero. 
Assuming $E_{0} = 0$ in (57), we obtain
\begin{equation}
\label{eq57}
\frac{{d^{2}\delta \xi} }{{dt^{2}}} = - \kappa \cdot \frac{{3}}{{2}} \cdot 
cH.
\end{equation}

To make formula (\ref{eq57}) consistent with formula (\ref{eq6}), it should be assumed that 
the coefficient $\kappa $ is close to unity. More precisely,
\begin{equation}
\label{eq58}
\left. {{\begin{array}{*{20}c}
 {if\quad \kappa = \frac{{2}}{{3}},} \hfill & {then} \hfill & 
{\frac{{d^{2}\delta \xi} }{{dt^{2}}} = - cH} \hfill \\
 {if\quad \kappa = 1,} \hfill & {then} \hfill & {\frac{{d^{2}\delta \xi 
}}{{dt^{2}}} = - \frac{{3}}{{2}} \cdot cH} \hfill \\
\end{array}} } \right\}.
\end{equation}

We arrive at the conclusion that by fitting one constant multiplier CGD 
allows the observed anomalous component of the acceleration of spaceships 
Pioneer 10, 11 to be described. The available experimental data on the 
spacecrafts Pioneer 10, 11 can be described by formula (\ref{eq57}) with $\kappa $ 
ranging from $\kappa = 2/3$ to $\kappa = 1$.

\subsection*{6. Conclusion}

\bigskip

From formulas (\ref{eq57}), (\ref{eq58}) it follows that as the spacecraft moves away from 
the Sun along radius: 

1. The additional acceleration is directed oppositely to the spacecraft 
direction and is constant in magnitude.

2. As long as the spacecraft can be viewed as a trial body, the acceleration 
is independent of the spacecraft characteristics and is of universal nature.

It should be emphasized that all the consideration is valid only when the 
initially made assumptions are fulfilled. Below they are mentioned in the 
explicit form.

\underline {First}, the space-time dynamics is described by the generalized 
equations of general relativity - conformal geometrodynamics equations (\ref{eq7}). 

\underline {Second}, the principal term in the expansion of $A_{0} $ is a 
constant in the space-time domain under consideration and the next term is a 
quantity of the fourth order of smallness. These assumptions are nontrivial, 
as in this manner cosmological-origin quantity $L = c/H$ is introduced to 
the construction of the $A_{0} $ to determine asymptotic expressions for the 
components of vector $A_{\alpha}  $ that satisfy equations (\ref{eq7}). In the 
general relativity there is no vector $A_{\alpha}  $, so $L$ cannot be 
introduced naturally in it. 

In the space-time domains, where the above two assumptions are not 
fulfilled, first, expression (\ref{eq39}) for$A_{0} $ can have a higher order of 
smallness, second, terms of another type (for example, pole terms) can 
become principal in the construction of $A_{0} $.

Expression (\ref{eq53}) for the additional acceleration of the trial body admits a 
direct experimental verification. Thus, it predicts:

- universal nature of the additional acceleration for all bodies moving 
along circular orbit [only the first term in the right-hand side of (\ref{eq53}) 
remains],

- possibility to change the acceleration direction in the motion along 
radius (in moving away - towards the Sun, in approaching - away from the 
Sun). 

In conclusion note that in equations (\ref{eq52}), (\ref{eq53}) the additional acceleration 
owes its appearance either to the terms including derivatives of $A_{\alpha 
} $ vector components or to terms that vanish in disappearance of this 
derivative type. The thermodynamic analysis performed, in particular, in 
ref. [7] suggests that these terms determine the geometrodynamic continuum 
viscous stress tensor. Therefore we conclude that the additional 
acceleration owes its origin to the geometrodynamic continuum viscosity.

The appearance of the anomalous component (provided that it is not explained 
by non-gravitational effects) raises a number of basic issues. For example, 
as applied to the general relativity, this is the issue of energy 
conservation law feasibility degree and cosmological expansion effect on the 
experiments within the Solar system. 

Spacecraft experiments can be undertaken in the future in order to arrive at 
answers to fundamental questions of the space-time theory. It is not 
improbable that the results of this paper will prove useful in development 
of programs of investigations in these experiments.

\bigskip

This work was carried out under financial support by the International 
Science and Technology Center (ISTC Project \#1655).

\bigskip

\subsection*{References}

\bigskip

[1] J. D. Anderson, P. A. Laing, E. L. Lau, A. S. Liu, M. M. Nieto, and S. 
G. Turyshev. Phys. Rev. \textbf{D. 65}, 082004/1-50 (2002). 
[arXiv:gr-qc/0104064].

\noindent
[2] Slava G. Turyshev, Michael Martin Nieto, and John D. Anderson. 
[arXiv:gr-qc/0503021].

\noindent
[3] M.V. Gorbatenko, A.V. Pushkin.\textit{ //} Voprosy Atomnoi Nauki i 
Tekhniki. Ser.: Teoreticheskaya i Prikladnaya Fizika. \textbf{2/2}, 40 
(1984).

\noindent
[4] Yu.A. Romanov.\textit{ //} Voprosy Atomnoi Nauki i Tekhniki. Ser.: 
Teoreticheskaya i Prikladnaya Fizika. \textbf{3}, 55 (1996).

\noindent
[5] A.Z. Petrov. New methods in the general relativity. Moscow, Nauka 
Publishers (1966).

\noindent
[6] M.V. Gorbatenko, A.V. Pushkin, H.-J. Schmidt. General Relativity and 
Gravitation.\textit{} \textbf{34}, No. 1. 9 (2002).

\noindent
[7] M.V. Gorbatenko. General Relativity and Gravitation.\textit{ 
}\textbf{37}, No. 1. 81 (2005). 

\noindent
[8] A. Einstein, L. Infeld, and B. Hoffmann. Ann. Math. \textbf{39.} 65 
(1938).

\noindent
[9] A. Einstein, L. Infeld. Can. J. Math. \textbf{1.} 209 (1949).

\noindent
[10] M.V. Gorbatenko, T.M. Gorbatenko. Theor. and Math. Physics. 
\textbf{140}. No.1. 1028 (2004).

\noindent
[11] M.V. Gorbatenko. Theor. and Math. Physics. \textbf{142}, No.1. 138 
(2005).

\end{document}